# Bromine-methanol etching of semiconductor crystals $Cd_{1-x}Zn_xTe_{1-y}Se_y$ with different selenium concentrations

*S.V. Naydenov*[*], *G.M. Babenko*, *O.K. Kapustnyk*, *I.M. Pritula*

Institute for Single Crystals of the National Academy of Sciences of Ukraine,
60 Nauky Ave., 61072 Kharkiv, Ukraine

The effect of the selenium concentration in the composition of the $Cd_{1-x}Zn_xTe_{1-y}Se_y$ semiconductor crystals on their etching with a bromine-methanol solution was studied. A thermodynamic model is proposed to describe the degree of etching of $Cd_{1-x}Zn_xTe$ and $Cd_{1-x}Zn_xTe_{1-y}Se_y$ crystals. A thermodynamic law is obtained for the first time to describe the change in the etching rate of $Cd_{1-x}Zn_xTe_{1-y}Se_y$ crystals with different selenium concentrations. Experimental dependencies are constructed for etching trajectories and etching rates with 5% bromine-methanol solutions of $Cd_{1-x}Zn_xTe_{1-y}Se_y$ crystal samples with nominal $x \approx 0.1$ and selenium concentrations of $y=0$, $y=0.02$, $y=0.06$, and $y=0.1$. The average etching rates were 24 μm/min, 18 μm/min, 15 μm/min, and 13 μm/min. The threshold effect of a strong decrease in the etching rate upon transition from ternary $Cd_{1-x}Zn_xTe$ to quaternary $Cd_{1-x}Zn_xTe_{1-y}Se_y$ crystals, associated with hardening of the crystal structure, is identify and theoretically explained. The obtained experimental data are in good agreement with the theoretical estimates and will be useful for choosing the optimal regimes of crystal treatment.

**Keywords**: Semiconductors II-VI, CdZnTeSe crystals, Etching, Bromine-methanol solution, Crystal characterization, Processing materials, Damaged layer.

## 1. Introduction

Semiconductor crystals of $Cd_{1-x}Zn_xTe_{1-y}Se_y$ solid solutions (CZTS) have appeared relatively recently [1, 2] and are considered new and promising materials for X-ray and gamma-ray radiation detectors [3]. The study of the structural [4], electronic [5, 6], electrical [7], detector [8] and other properties of CZTS crystals continues. The quality of detector elements made of these crystals depends largely on the conditions and methods of their processing. The cutting and polishing of AIIBVI semiconductor crystals almost always leads to the formation of a deep damaged layer near the surface with a thickness ranging from several tens of microns to one hundred microns [9, 10]. Defects formed inside the damaged layer block the collection of charge carriers from the material volume. They also cause surface leakage currents. Therefore, without removing the damaged layer, the samples do not exhibit detector properties. Various methods have been used to remove the damaged layer of CZTS crystals. Many of these methods are similar to the methods used to treat more studied ternary $Cd_{1-x}Zn_xTe$ (CZT) crystals. First, bromine is chemically etched (polished) with bromine solutions of different concentrations [11-13]. To date, there are practically no data on the chemical and mechanical polishing of CZTS crystals in the scientific literature. The choice of the optimal mode of crystal processing (etchant composition, etching temperature and time, other conditions) is one of the primary tasks in the fabrication of semiconductor detectors.

Theoretical and experimental studies indicate a significant difference between CZTS and CZT materials. The addition of selenium changes the thermodynamic properties of the solid solution; starting from a certain threshold, the excess Gibbs energy decreases. This affects the stability of the crystal structure of the CZTS crystals: the number of extended defects, tellurium inclusions, clusters of dislocations, etc., decreases. Theoretical analysis [14] revealed that the crystal structure of CZTS crystals increases with increasing selenium concentration. This effect was confirmed qualitatively by comparing the crystal properties of CZTS. With increasing selenium concentration, there is an apparent decrease in the number of some characteristic extended defects in these crystals. However, there has been no direct quantitative confirmation thus far. In this work, the chemical etching (polishing) of the surface of CZTS crystal samples with different selenium contents — $y = 0$ (without selenium), $y = 0.02$, $y = 0.06$ and $y = 0.1$ in the anionic sublattice — was studied. A 5% solution of bromine in methanol was used as a polishing etchant. The change in surface morphology at different etching stages was investigated. Characteristic "etching trajectories" (time dependence of etching depth) were constructed, and average etching rates were determined. The influences of temperature, stirring and other factors on the etching rate were considered. Even at a small selenium concentration of approximately $y \sim 0.02$, there was a sharp decrease of 20% in the etching rate of the CZTS quaternary crystals compared with that of the CZT ternary crystals. This effect was detected for the first time and provides direct evidence for the hardening of the CZTS crystal matrix. This phenomenon is also well explained within the framework of the proposed simple thermodynamic model of crystal etching. In addition, the features of chemical etching of CZTS crystals established here are important for choosing the correct methodology for the fabrication of new semiconductor detectors.

---

[*] sergei.naydenov@gmail.com

## 2. *Theoretical background*

Cutting a semiconductor crystal leads to the formation of a damaged layer near the cutting surface. Its thickness depends both on the material itself and on various external factors. For example, the choice of cutting method and cutting material, depth, cutting force and speed, etc. CdTe, CZT, and CZTS crystals, as well as other crystals of the AIIBVI group, are "soft-brittle" materials [15]. They have relatively low hardness (~2–3 Mohs scale for scratching) and, at the same time, are rather brittle materials. Even small external loads applied to these crystals can lead to plastic deformation and severe surface damage (scratches, acquired defects). Diamond cutting machines are most often used to cut these crystals. The diamond thread has a diameter of one hundred microns or more, and the thickness of the "diamond" coating itself is tens of microns. Therefore, such cutting produces an even but often too wide cut. The thickness of the damaged layer after such cutting, taking into account the entire area adjacent to the cutting surface with plastic deformation, is estimated to be approximately 100 microns [9]. Mechanical grinding and finishing (mirror) polishing reduce the thickness of the damaged layer to 50 microns [10]. In the case of particularly thorough polishing, the thickness of the damaged layer can be several times less.

To remove the damaged layer of CZT and CZTS crystals by chemical polishing, solutions based on different active "polished" substances, such as bromine, iodine, and other strong oxidizing agents, can be used [16-18]. One of the standard etchants is a bromine-methanol solution of $Br_2$ (bromine) in $CH_3OH$ (methanol). Let us consider the chemical reactions that occur during such etching. Assuming that methanol does not enter any reactions, we have the following general equation:

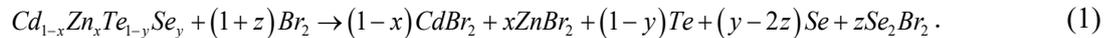

$$Cd_{1-x}Zn_xTe_{1-y}Se_y + (1+z)Br_2 \rightarrow (1-x)CdBr_2 + xZnBr_2 + (1-y)Te + (y-2z)Se + zSe_2Br_2. \quad (1)$$

In this composite reaction, at room temperature, the released atomic tellurium does not react with bromine. For selenium, such a reaction is possible with the formation of some amount of diselenium dibromide. The reaction to form selenium tetrabromide is unlikely because $SeBr_4$ has more than twice the standard enthalpy of formation (-70 kJ/mol) compared with $Se_2Br_2$ (-31 kJ/mol). In this case, we assume that the activation energy for the formation of the compound complex $SeBr_4$, containing four bromine atoms, is several times greater than the activation energy of the complex $Se_2Br_2$, in which there is only one bromine atom per selenium atom. As a result, the rate (probability) of the reaction (1) of selenium tetrabromide formation, according to the Arrhenius law, will be an order of magnitude lower and will not strongly affect the course of the etching reaction within the characteristic etching times. Note that selenium derivatives in the proposed reaction have virtually no effect on further analysis and results, since these reaction products will in any case be small due to the smallness of the initial concentration of selenium in the solid solution (the smallness of the parameter $y \ll 1$).

Cadmium and zinc bromides have high solubilities in water and organic solvents. Therefore, they are well washed from the crystal surface during the etching process. In contrast, tellurium and selenium are poorly soluble and can accumulate on the surface and/or be adsorbed from the solution during the etching process. This leads not only to non-stoichiometry of the initial material surface but also to the formation of additional defects on it (inclusions, micro-formations, and pores). At long etching times (more than 3–5 min in 5% bromine-methanol solution), the surface cleanliness and smoothness (roughness), as a rule, begin to deteriorate.

Let us compare the etching reactions of CZTS and CZT crystals (both with the same zinc concentration) from a thermodynamic point of view. For this purpose, let us consider a formal closed system of "etching

"comparison" (a mental experiment), in which two ideal samples, CZTS and CZT, are simultaneously etched in the same etchant under conditions of thermodynamic equilibrium. This is acceptable if the etching is fast enough and all nonequilibrium fluxes disappear. The inverse of etching is the process of the adsorption of atoms on the surface. The ideality of the samples implies the absence of any disturbances on their surface. For example, they can be crystals obtained by chipping or natural single crystals. It is convenient to assume that the part of the starting material that is vented and enters the liquid solution is an additional component of the system. For simplicity, we assume that both solid solutions and etching solutions are ideal [19]. Let us consider that the mole fraction of selenium in CZTS crystals is small. Therefore, the mole fractions of etched substances for CZTS and CZT are not significantly different from each other. This allows us to neglect additional contributions in the Gibbs energies for the subsystems and write the condition of phase equilibrium (equality of average molar Gibbs energies in different phases) in the form of

$$\mu_{CZTS}^0 + RT \ln(1 - w_{CZTS}) \approx \mu_{CZT}^0 + RT \ln(1 - w_{CZT}), \tag{2}$$

where $\mu_{CZTS}^0$ and $\mu_{CZT}^0$ are the chemical potentials of "pure", i.e., unetched crystals; $R$ is the gas constant; $T$ is the absolute temperature; $w_{CZTS}$ and $w_{CZT}$ are the mole or weight fractions of the etched substance, i.e., the "solubility" of the starting material in the etchant. Let only a small fraction of the starting material pass into the liquid solution during etching, i.e., $w_{CZTS} \ll 1$ and $w_{CZT} \ll 1$. Then, instead of the ratio, we obtain

$$w_{CZTS} - w_{CZT} \approx \frac{\mu_{CZTS}^0 - \mu_{CZT}^0}{RT} = -\frac{|\Delta G_{ex}|}{RT}, \tag{3}$$

where $\Delta G_{ex} \leq 0$ is the negative excess Gibbs energy of formation of the CZTS solid solution with respect to the CZT solid solution calculated in [14]. The mole fraction of etched matter for the CZTS crystal should be smaller than that for the CZT crystal. Note that the formula does not impose any restrictions on the values of $w_{CZTS}$ and $w_{CZT}$. They can remain arbitrary but small enough to fulfill the approximations of the proposed model.

Let us assume that etching is homogeneous over the entire surface of the samples. Then, the mole fraction $w_*$ and the thickness $\Delta L_*$ of the layer of substance etched from the entire surface of the sample, under the condition of thermodynamic equilibrium in the system, are related by a linear expression

$$w_* = \Delta L_* \left( \frac{S_{tot}}{V} \right), \tag{4}$$

where $S_{tot}$ and $V$ are the total surface area and volume of the sample, respectively. The ratio $V/S_{tot} = L_*$ specifies the characteristic linear size of the sample. For example, if the sample has the shape of a cube with edge $L$, then $L_* = L$. Furthermore, we assume that the linear dimensions of the samples in the etching system introduced in this way coincide, i.e., $L_{CZTS} = L_{CZT}$. If these dimensions do not coincide, the conditions of thermodynamic equilibrium in the general system are formally violated, and etching of one of the samples will occur faster than the other, and exactly until the specified dimensions coincide.

The expression was obtained in the thermodynamic limit for a closed system. Let the relaxation time, during which equilibrium in the system is reached, be equal to $\tau_*$. During this time, the thickness of the sample decreases by a characteristic value $\Delta L_*$. Suppose that the etching conditions are such that this substance is entirely transferred into the liquid etching solution, which feeds by the etchant so that its concentration does not change, i.e., let us move from a closed to an open system. Then, for the next characteristic time interval $\Delta t = \tau_*$, the thickness of the sample decreases once again by the value $\Delta L_*$. For some finite time of real etching, the thickness of the sample will change $n = t/\tau_*$ times. The total thickness change in such a time-homogeneous etching is equal to $\Delta L(t) = n \Delta L_*$. Thus, instead of equilibrium, we have a stationary etching process. Let us introduce the rate of real etching $V(t)$ via a natural definition

$$V(t) \equiv \frac{\Delta L(t)}{\Delta t}. \tag{5}$$

When comparing real etching with the considered model system, we have

$$\frac{\Delta L(t)}{\Delta t} \approx \frac{\Delta L_*}{\tau_*}. \quad (6)$$

The approximate equality in the formula takes into account possible deviations from the ideal course of etching. From the formula, the thermodynamic *law of etching* is as follows:

$$V_{CZTS} - V_{CZT} = -V_* \frac{|\Delta G_{ex}|}{RT}, \quad (7)$$

where $V_* = L_* \tau_*^{-1}$ is a characteristic dimension *constant of etching*, which has the dimension of the (etching) rate and depends on the mutual properties of the etching system of two crystals. Strictly speaking, for unambiguous determination of $V_*$ and correct comparison of real etching rates of a pair of CZTS and CZT crystals, the etching conditions (type and concentration of etchant, etching temperature, nature of solution mixing, other parameters) should be the same. This corresponds to the stationarity condition in the proposed theoretical model. Note that the introduced value of $V_* = V_*(x, y)$ may depend on the composition $(x, y)$ of the crystals in the etching system. This is because the establishment of a stationary state during the etching of crystals with significantly different compositions may require more time $\tau_*$ than for crystals with a similar composition.

In accordance with the obtained formula, the etching rate of CZTS crystals should always be less than the etching rate of CZT crystals (under the same etching conditions). Thus, with increasing temperature, this difference should disappear. In addition, the change in the value of $|\Delta G_{ex}|$ with a change in the selenium concentration in the CZTS crystal should correlate with the corresponding change in the CZTS etching rate compared with the CZT etching rate. This correlation takes place in the experiment with etching in a bromine-methanol solution (Section 4).

Thus, there is a fundamental difference in the etching rates of CZTS and CZT materials, and at low concentrations, these solid solutions are proportional to the excess Gibbs energy of formation of the quaternary CZTS solid solution instead of the ternary CZT solid solution. Weak effects of orientation dependence, when the etching rate on different crystallographic planes (faces) of the crystal may slightly differ, are not considered here. The presence of a damaged layer, i.e., the nonideality of the crystal, also does not violate the above reasoning, since changes in all thermodynamic quantities on the surface are always small compared with their changes in volume. New peculiarities arise only if the crystal surface is treated differently at different sites. For example, some areas are not processed at all or are only roughly polished, and others have finished polishing, etc. With such a heterogeneous nature of surface treatment, the conditions of "thermodynamic equilibrium" for the etching system will initially be violated. Therefore, etching rates and/or their characteristic differences when comparing etching of different crystals may differ significantly in areas with different types of processing.

## 3. Materials preparation

CZTS crystals grown at the Institute for Single Crystals of the National Academy of Sciences of Ukraine via the vertical Bridgman method under high argon pressure were used for the experiments [20]. All the crystals were obtained under close growth conditions and were similar to each other with respect to their crystalline (structural) perfection. The phase composition of the crystals was confirmed via powder X-ray diffractometry and electron spectroscopy (on a low-vacuum scanning electron microscope with a low-temperature characteristic X-ray detector). Samples with a typical size of 8×8×5 mm$^3$, cut in parallel shapes from the middle of $Cd_{1-x}Zn_xTe_{1-y}Se_y$ crystal ingots with the same zinc content of approximately $x \approx 0.1$ and different selenium contents of $y = 0$, $y = 0.02$, $y = 0.06$, $y = 0.1$, were selected for etching. The sample with $y = 0$ corresponds to the CZT crystal without the addition of selenium. Before etching, all sides of the samples were successively ground with boron carbide powder with fraction sizes of 9 and 3 microns. The damaged layer on such faces reaches up to one hundred microns. In addition, one of the pairs of opposite "working" faces was subjected to finish polishing with diamond powder. The resulting average roughness of the base surface is $R_a$ = 5–12 nm. However, deep scratches with depths of several microns may still be present on the facets. The damaged layer on the "working" faces is several tens of microns thick. When manufacturing detector elements, electrical (most often ohmic) contacts are applied to the finished (mirror polished) faces.

To eliminate the damaged layer remaining after mechanical polishing of CZT and CZTS crystals, chemical polishing with bromine-containing solutions is most often used, particularly chemical polishing in solutions of molecular bromine Br2 in methanol. In this work, 5% bromine-methanol solution was chosen as the polishing etchant. This solution is convenient for processing and investigating the properties of CZTS crystals. An etchant with a lower bromine concentration is not convenient for a sufficiently accurate determination of the etching trajectory (time dynamics). With decreasing bromine content, the etching rate decreases significantly, which, at moderate temperatures ranging from +18°C to +25°C, is usually directly proportional to the bromine concentration in the solution. As a result, the change in the thickness of the material over a characteristic time of approximately 30 s, i.e., the minimum etching depth, becomes too small (only a few microns); therefore, the errors in its determination (using a micrometer) increase dramatically. In contrast, at very high bromine concentrations, the etching rate of the crystal is too high. In this case, the quality of chemical polishing deteriorates due to rapid degradation and strong surface etching, especially at long processing times.

Our experiments revealed that the dynamics and results of CZTS etching strongly depend on the quality and freshness of the bromine-methanol solution used. After several hours of overexposure to the etching solution, the observed crystal etching parameters changed significantly. Therefore, only freshly prepared bromine-methanol solutions obtained from chemically pure components were used for sample processing. In addition, between the intermediate etching steps, the samples were washed with methanol and/or deionized water to remove the residues of the used solution and any released or background substances. This is necessary to maintain cleanliness and the same (external) etching conditions throughout all the etching steps, as any contaminants from the etching solution or crystal surface can inhibit the etching processes. The use of distilled water for washing does not guarantee the complete purity of the prepared surface; after etching, many centers with broken chemical bonds (mainly tellurium) remain, to which various ions of contaminants from the external environment can actively adsorb.

## 4. *Experimental part*

A series of experiments on chemical polishing (etching in 5% bromine-methanol solution) of crystalline samples of $Cd_{1-x}Zn_xTe_{1-y}Se_y$ with the same zinc concentration $x \approx 0.10$-$0.11$ and significantly different selenium concentrations $y$ were performed (see Figs. 1-3). Composition and dimensions of the studied samples: CZTS-10-5-1 ($x = 0.105$, $y = 0$), $5.3 \times 7.8 \times 8.3$ mm$^3$; CZTS-9-4-1 ($x = 0.114$, $y = 0.022$), $5.5 \times 7.6 \times 8.1$ mm$^3$; CZTS-11-3-2 ($x = 0.109$, $y = 0.06$), $6.5 \times 7.1 \times 7.3$ mm$^3$; CZTS-12-4-2 ($x = 0.104$, $y = 0.1$), $6.4 \times 7.1 \times 7.5$ mm$^3$. These samples are hereafter designated CZT, CZTS2, CZTS6, and CZTS10 in accordance with the average (nominal) selenium content of the crystalline ingots from which they were made. All the samples were ground after being cut. The pair of opposite faces with the smallest distance between them (the first digit in the sample dimensions) was mirror polished. Before etching, the sample surfaces were cleaned in an ultrasonic bath to remove traces of inorganic and organic contaminants, as well as residual polishing compounds remaining after crystal processing. Fig. 1 shows photos of typical polished surfaces of samples before etching.

The etching of each sample was carried out with freshly prepared 5% bromine-methanol solutions separately in fluoroplastic containers. However, this process was carried out in the same way for all samples. The etching consisted of several stages with durations ranging from 30 s to 2.5 min. The chosen *etching formula (main stages)* was as follows: 30 s + 30 s + 30 s + 30 s + 30 s + 30 s + 30 s + 2.5 min + 2.5 min + 2.5 min + 2.5 min. The total duration of the etching was up to 8 min. Between the individual stages, the samples were washed, and all their dimensions were measured. Care was taken to minimize the samples exposure to air and to prevent interaction with the ambient atmosphere. Before etching, the prepared solution was kept for some time to stabilize the chemical processes and to ensure that its temperature matched that of the thermostat. The etching temperature was kept as constant as possible and equal to +19°C, with possible deviations not exceeding ±0.5°C. Fig. 2 shows photos of the polished surfaces of the samples after etching.

As an etching method, we used the conventional method of complete immersion of the sample in the etching solution. In this case, all sides of the sample of an initially regular rectangular shape (parallelepiped) with three pairs of parallel opposite sides are equally subjected to etching. This method has proven itself well in practice and is most often used in the technical processing of detector elements, which are made of CZT or CZTS semiconductor crystals. During etching, the sample was positioned at an angle to the vertical (in the edgewise state) and was rotated from time to time to ensure a more uniform supply of etchant to all side faces. The sample never contacted the bottom of the fluoroplastic etching container with its entire "lower" surface. Due to this, a possible difference in the etching character of the conventionally "upper" and "lower" faces was eliminated. Stirring of the solution during etching was not used, since it was found that it is an additional factor that significantly changes the etching rate. The uniformity of etching and the reproducibility of the etching results were controlled by repeating series of measurements on different samples made from the same crystal. In the absence of mixing, the experiments yield identical results within the accuracy of the sample thickness measurements. Experiments on chemical-mechanical etching of new crystal samples under reproducible hydrodynamic conditions (with controlled rate of mixing of the etching solution) are planned for the future.

The depth of etching was determined by the change in the geometric dimensions of the samples fabricated in the shape of a rectangular parallepiped. Care was taken to maintain the parallelism of the opposite faces of the original samples after polishing. During the measurements, it was assumed that etching of a pair of opposite sides occurred in the same way so that the depth of etching was equal to

$$\Delta L_i(k) \approx \frac{L_i(k+1) - L_i(k)}{2}, \qquad (8)$$

where $k$ is the number of etching stages (here $k = 0$ corresponds to the beginning of etching) and where $L_i(k)$ is the transverse thickness of the sample between a pair of opposite sides, which are numbered by the index $i = 1, 2, 3$ $\Delta L_i(k)$, is the depth of etching for the selected pair of sides. Deviations from parallelism of the faces of the original sample and spatial inhomogeneity of the etch result in thickness variations. To statistically minimize these errors, measurements were taken several times at the same fixed location of the selected face or averaged by measuring the transverse thickness at several different locations (at the corners and in the center of the face). A precision micrometer TESA® MICROMASTER with a division value of 1 μm was used to measure the dimensions. The error of the obtained experimental data corresponds to its maximum error, which, according to the passport, is equal to ± 4 μm. When the etching time and/or rate are too small, the measurement results may be unreliable. Therefore, we selected an etching mode in which the etching depth at any stage always exceeded 5–7 microns.

Approximation (8) is performed with good accuracy provided that pairs of opposite sides of the sample are etched equally. To achieve this, several additional techniques were employed: 1) the selected sides of the sample were always positioned vertically or at an angle to the vertical and were not horizontal; 2) the sample was never placed entirely by either side on the etchant container bottom; 3) the sample was slow rotated during etching, occasionally changing the direction of rotation, achieving slight mixing of the solution. Taking these techniques into account, we verified that the relative difference in the etching rate of opposite pairs of sample sides. This experimental test involved comparative etching of the same sample, first fully immersed in the solution, and then immersed in the etching solution on only one of the two selected sides (the other opposite side, either left out of the solution or protected from etching by a coating). In the first step, the total etching depth of both sides is divided in half. The difference in measurements of etching rates practically does not exceed 5-10% over etching time intervals of several minutes. Therefore, in the used immersion etching method, the difference in the etching depth of opposite sides of the samples remains equal within the accuracy of their thickness measurements with a micrometer.

Using the measured data, we can plot the *etching trajectory* $T_i(k) = \Delta L_i(1) + \ldots + \Delta L_i(k)$ and determine the *average etching rate* at each stage via the formula $V_i(k) = \Delta L_i(k)/\Delta t_k$, where $\Delta t_k$ denotes the duration of the $k$ stage. The determination of the etching rate is more formal since, in general, it depends in a complex way on the microscopic state of the crystal surface, which changes significantly during prolonged etching, despite the maintenance of the same external conditions. The etching of the surface of a real crystal and the change in the true (thermodynamically equilibrium) etching rate may depend on the internal crystal structure, the presence and distribution of extended defects that reach the surface during the etching process, etc. However, in general, the etching dynamics are fairly well described by the pair of experimental values $T_i(k)$ and $V_i(k)$. As we believe, this is due to the self-averaging effect of microscopic variations in the etching rate on different inhomogeneities of the crystal for characteristic times exceeding several tens of seconds. This general conclusion is well confirmed by etching experiments using different etching formulas for "similar" samples made from the same crystal parts. Fig. 3 shows typical photos of the surface areas of the samples at the end of etching, which were obtained under an optical microscope.

The concept of the "etching trajectory" that we introduced has a simple physical meaning – it is the dependence of the total etching thickness on the total etching time at all successive stages. In the mathematical sense, the etching trajectory is an integral characteristic that is obtained by formal integration of the instantaneous etching rate. It is important to note that with such integration, measurement errors are actually reduced – measurement variations are smoothed out. In contrast, the etching rate is a differential characteristic of the process; mathematically, it is a derivative quantity. And, like any derivative, it can change dramatically over time. The more so, the shorter the measurement interval. Its variations depend much more on various factors, which is observed in the experiments. During measurements, this value can accumulate errors and the accuracy of its determination with a strong decrease in the etching time becomes too low. In contrast, the average etching rate will be a well-defined characteristic of the process. Taking these features into account, we choose the etching trajectory and the average etching rate (at certain etching stages or along the entire etching trajectory) to control the etching processes.

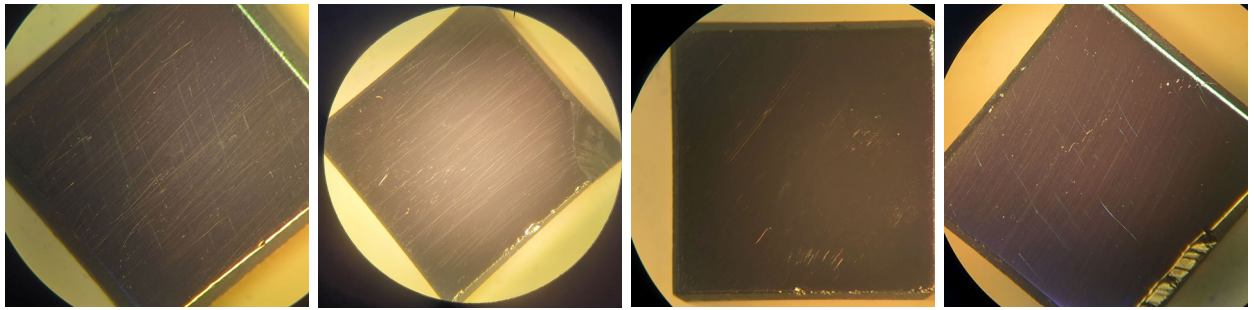

**Fig. 1.** Images of the polished faces of the crystal samples (from left to right) CZT (a), CZTS2 (b), CZTS6 (c), and CZTS10 (d) before etching.

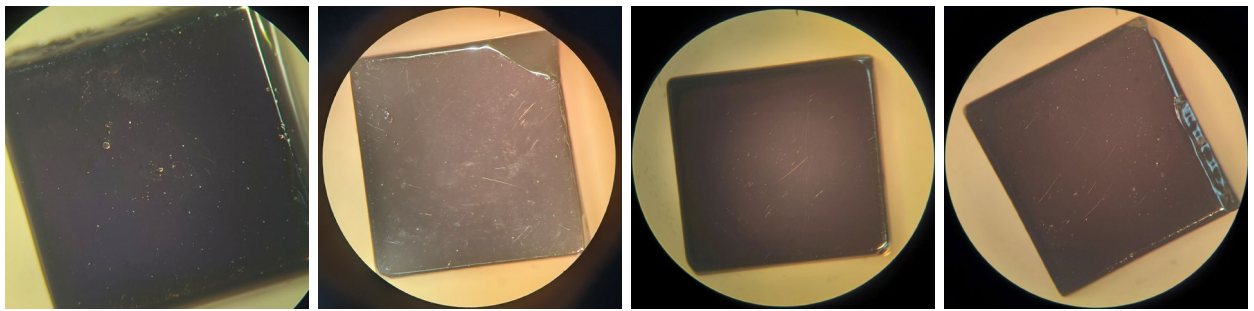

**Fig. 2.** Images of the polished faces of the crystal samples (from left to right) CZT (a), CZTS2 (b), CZTS6 (c), and CZTS10 (d) after etching.

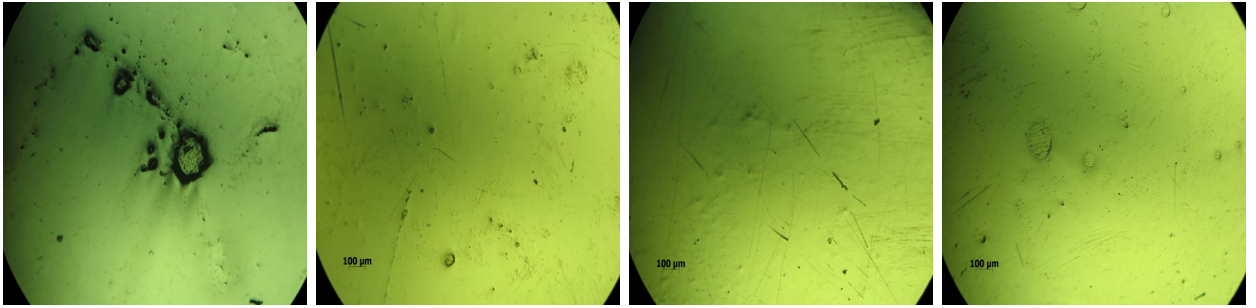

**Fig. 3**. Microscopy images (x100) of the polished faces of the crystal samples (from left to right) CZT (a), CZTS2 (b), CZTS6 (c), and CZTS10 (d) after 8 min of etching.

## 5. *Results and discussion*

Fig. 4 shows the etching trajectories of the CZT, CZTS2, CZTS6, and CZTS10 samples with a total etching time of 8 min. The dots correspond to the experimental values. With increasing selenium concentration in the samples, the angular slope of the plotted curves, i.e., the etching rate of the crystals, decreases. The general dependence of the etching trajectory at the initial stages is close to linear, but in general, the course of etching differs. This may be due to the different degrees of initial mechanical polishing of the surface (it may be rougher, and its initial etching is usually faster), as well as the role of agitation in the etching of some samples. In the absence of agitation, etching can be slowed if the etching products inhibit the reaction. This can be clearly seen in the etching of the control sample CZT, which was etched without stirring. Separate experiments involving the etching of samples with similar compositions and similar degrees of mechanical polishing revealed that stirring can increase the etching rate by 1.5-2 times. For this reason, for example, the initial etching step of CZTS2 occurs at a slightly excessive etch rate. In addition, as the damaged surface layer – estimated to be approximately 30–50 μm thick – is progressively removed, the etch rate gradually decreases with increasing etching time. The reason for this is that the bulk crystalline structure of the sample, which is much stronger than the surface damaged layer, begins to undergo etching.

The etching of the CZTS crystal surface with 5% bromine-methanol solution, as a rule, results in nonuniform characteristics. Areas of the surface with initial growth defects or traces of mechanical processing undergo deeper etching. The polishing effect is observed at etching times exceeding 30 seconds (see Fig. 2). At short etching times, the effect of chemical polishing is more pronounced, and the surface roughness is significantly reduced. However, as a result of long etching times (more than 3–5 minutes in total), the roughness, on the contrary, can increase. This behavior resembles the features of CZT crystal etching [11, 21] and is the subject of a separate study. At very prolonged etching (more than 5 min), the corners of samples or places with large damage (chips, cracks, nicks, crystal grain boundaries, etc.) are strongly stratified. In the vicinity of such macro-uniformities, strong concentration gradients of the etching composition appear, and etching occurs much faster. To improve the uniformity of etching, the solution can be stirred or the samples can be rotated (rocked), which results in the same results. However, as already mentioned, stirring leads to a strong increase in the etching rate, which is not desirable. The peculiarities noted here should be considered in the practical processing of CZTS crystals and the fabrication of detector elements from them.

For the obtained etch trajectories, the etch rates at different stages were calculated. These data are presented in Table 1. The etch rate at stages of short duration has significant variations. This is especially the case at the initial stages of etching. This is due to measurement errors in the thickness of the removed layer, as well as to the unevenness of the etching of surfaces with different degrees of roughness. When the average etching rate is determined for sufficiently long times (several minutes or more), these variations are reduced. The average etching rates for the full etch paths of the CZT, CZTS2, CZTS6, and CZTS10 samples were 24 μm/min, 18 μm/min, 15 μm/min, and 13 μm/min, respectively. With increasing selenium content in the CZTS crystals, this rate decreased significantly. This conclusion is also confirmed qualitatively by analyzing the surface morphology obtained after prolonged etching. A comparison of the microscope images shown in Fig. 3 reveals that the CZT surface is etched near natural defects more strongly than the other cases are. Moreover, the visible traces of etching near growth defects in the CZTS crystals became increasingly indistinct with increasing selenium concentration.

The observed decrease in the etching rate of the CZTS crystals with increasing selenium concentration is related to the hardening of the crystal lattice when selenium is introduced into the solid solution matrix. In fact, the decrease in the rate of nonselective (polishing) etching is direct evidence of the hardening of the crystal structure of this material.

Let us compare the etching results with the theoretical law. At low selenium concentrations, the thermodynamic rate of $V_*$ should not differ much from the etching rate of the CZT crystal (without selenium). However, etching affect both pairs of opposite faces of the sample. Therefore, the total depth of etching $\Delta L_*$, the mole fraction $w_*$ of etching and the $V_*$ rate itself should double, i.e., $V_* \approx 2V_{CZT}$. With this in mind, we obtain the relation

$$\delta V/V \equiv 1 - \frac{V_{CZTS}}{V_{CZT}} \approx \frac{2|\Delta G_{ex}|}{RT} \qquad (9)$$

for the relative change in the etch rate as a function of the (small) selenium concentration in the CZTS crystal. Calculation of the excess Gibbs energy $|\Delta G_{ex}|$ for a selenium concentration of $y = 0.02$ gives a value of 260 J/mol [14]. Substituting this value into the formula, we obtain an estimate $(\delta V/V)_{theory} \sim 22\%$ for room temperature. A comparison with the experimental values of $(\delta V/V)_{exp} \sim 24\%$ shows good agreement between the theoretical and experimental results. Note the strong (threshold) effect of changing the etching rate. This is enough to increase the selenium content in the CZTS crystal by only a few atomic percentages and leads to a change in the etching rate of several tens of percent, i.e., an order of magnitude greater.

With increasing selenium concentration in the CZTS samples, their etching rate decreases but is not as fast as that at low selenium concentrations. There is a gradual equalization of etching rates. This can be clearly seen by comparing the etching trajectories in Fig. 4. There is a strong difference in the etching trajectories of the CZT and CZTS2 crystals, whose selenium contents are low and differ by $y = 0.02$. The divergence of the etch trajectories decreases significantly for CZTS2 and CZTS6 crystals with higher selenium contents, although this difference itself becomes twice as large as that in the previous case. Finally, for the CZTS6 and CZTS10 crystals (the difference in the selenium content is also $\Delta y = 0.04$), the etch trajectories become even closer together. This pattern reflects the threshold character of the hardening of the crystal lattice when selenium is added to it. The effect of such hardening begins at a certain critical concentration of selenium (in [14], it is estimated by the value $y \approx 0.18$ at a fixed zinc concentration of $x = 0.1$). The degree of hardening increases with increasing selenium concentration $y$ since the excess Gibbs energy, which has negative values, monotonically increases in absolute value $|\Delta G_{ex}(y)|$ in the framework of the weakly dilute solid solution model. However, the steady-state etching rate $V_*(y)$ due to the hardening decreases even faster (by a factor of approximately ten) with increasing selenium concentration. As a result of this compensation, a "convergence" of etching rates of CZTS crystals of different compositions is observed when $\delta V(y_1) \approx \delta V(y_2)$ at high selenium concentrations $y_1 \neq y_2$. At very high selenium concentrations in CZTS crystals of different compositions, one should not expect a significant difference in the rates of etching by bromine-methanol solutions.

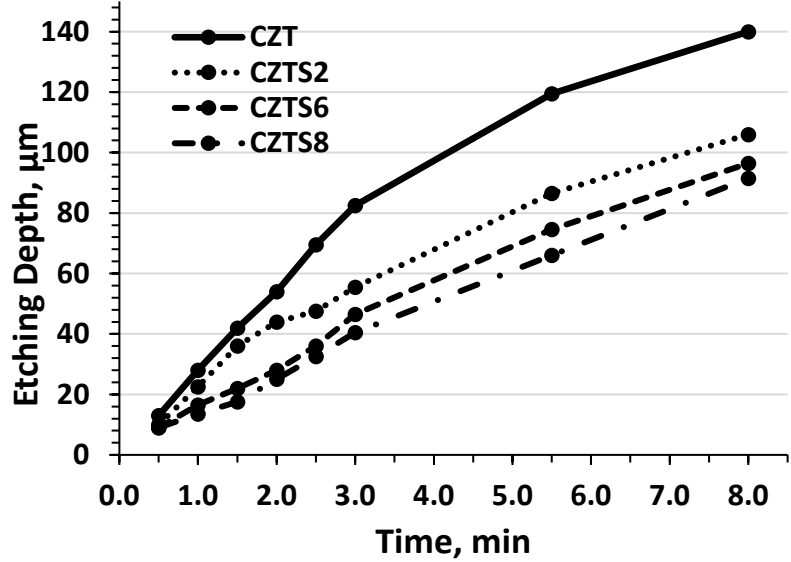

**Fig. 4.** Etching trajectories of samples CZT, CZTS2, CZTS6, and CZTS10.

**Table 1.** Etching rate (μm/min) at different stages for the CZT, CZTS2, CZTS6, and CZTS10 samples and the average speed of total etching (the last row of the Table).

| Etching stage | $V_{CZT}$ | $V_{CZTS2}$ | $V_{CZTS6}$ | $V_{CZTS10}$ |
|---|---|---|---|---|
| 0.0-0.5 min | 26 | 20 | 17 | 17 |
| 0.5-1.0 min | 30 | 25 | 16 | 15 |
| 1.0-1.5 min | 28 | 21 | 15 | 13 |
| 1.5-2.0 min | 24 | 16 | 14 | 13 |
| 2.0-2.5 min | 31 | 23 | 16 | 11 |
| 2.5-3.0 min | 26 | 19 | 18 | 16 |
| 3.0-5.5 min | 15 | 12 | 11 | 10 |
| 5.5-8.0 min | 14 | 11 | 10 | 9 |
| 0.0-8.0 min (total) | **24** | **18** | **15** | **13** |

Interestingly, zinc and selenium play different roles in changing the properties of CZTS crystals. An increase in the zinc concentration leads to an increase in the etching rate. This phenomenon has been reported for CZT crystals when etched with solutions of strong acids [18, 22]. We have not performed etching experiments on CZTS crystals with the same selenium concentration but different zinc concentrations. However, we expect similar behavior for them as well. In other words, zinc addition leads to a weakening of the strength of the cationic sublattice and the crystal structure as a whole and thus to enhanced formation of various growth defects in such crystals. In contrast, the addition of selenium leads to a strengthening of the CZTS crystal structure, which is characterized by a sharp decrease in the etching rate and a reduction in the number of expected defects.

## 6. *Conclusions*

The addition of selenium to CZTS quaternary crystals significantly changes their crystalline properties and leads to hardening of the crystal structure. This effect has thermodynamic and chemical causes. There is a decrease in the excess energy of Gibbs during the formation of a solid solution containing added selenium. The covalent bonds in the cationic and anionic sublattices are also strengthened. The macroscopic manifestation and proof of the effect of hardening of the crystal structure is a sharp decrease in the rate of chemical etching of the CZTS crystals by bromine-methanol polishing solution compared with that of the CZT crystals. The nature and result of such etching of CZTS depend on the roughness of the initial surface, the presence and depth of the damaged layer after treatment of the crystal, the etching temperature and the mixing of the etching solution. The change in the etching rate of the CZTS samples with a small change in the selenium concentration is well described by the proposed linear thermodynamic law (the decrease in the etching rate is proportional to the increase in the absolute value of the excess Gibbs energy). At high selenium concentrations, the difference between the etching rates of CZTS crystals of different compositions gradually levelled off. The theoretical and experimental results obtained in this work can be used to select the optimal modes of processing and chemical polishing of detector elements made of CZTS crystals.